\documentclass[aps,pre,twocolumn,floats]{revtex4}

\usepackage{graphicx}  % standard LaTeX graphics tool
\usepackage{latexsym}   % useful for coding complex math
\begin{document}
\renewcommand{\thesection}{\arabic{section}}
\renewcommand{\thetable}{\arabic{table}}
\newcommand{\beq}{\begin{equation}}
\newcommand{\eeq}{\end{equation}}

\title{\bf Unveiling community structures in weighted networks}
\author{Nelson A. Alves\footnote{E-mail: alves@ffclrp.usp.br} }
\address{{\it Departamento de F\'{\i}sica e Matem\'{a}tica, 
        FFCLRP Universidade de S\~{a}o Paulo, Avenida Bandeirantes 3900,}\\
        CEP 14040-901, \thinspace\ Ribeir\~{a}o Preto, S\~ao Paulo, Brazil}
\date{February 27, 2007}

\vskip1cm 
\begin{abstract}
  Random walks on simple graphs in connection with electrical resistor networks 
lead to the definition of Markov chains with transition probability matrix 
in terms of electrical conductances.
  We extend this definition to an effective transition matrix $P_{ij}$ to 
account for the probability of going from vertex $i$ to any vertex $j$ of the 
original connected graph $G$. 
  Also, we present an algorithm based on the definition of this effective transition 
matrix among vertices in the network to extract a topological feature related
to the manner graph $G$ has been organized.
  This topological feature corresponds to the communities in the graph.
\end{abstract}

\maketitle

\vskip0.1cm {\it Keywords:} communities in networks, weighted graph, electrical network, 
          random walks, Laplacian spectrum.
 
\vskip0.1cm {\it PACS-No.: 89.75.-k, 89.75.Hc, 05.10.-a}
%%%%%%%%%%%%%%%%%%%%%%%%%%%%%%%%%%%%%%%%%%%%%%%%%%%%%%%%%%%

\vskip1cm 

\section{Introduction}
\indent
  Network modeling is becoming an essential tool to study and understand the 
complexity of many natural and artificial systems \cite{Barabasi05}.
  Applications \cite{BarabasiRev,MendesRev,NewmanRev} include technological networks 
as the Internet, World Wide Web and electric power grid; biological networks as
metabolic \cite{Barabasi02,Holme03,AmaralNat05}
and amino acid residue networks \cite{Vendruscolo,Greene03,Atilgan05,AlvesA07};
and far more studied, social networks.   
  This understanding firstly passes through the analysis of their topological features,
usually related to complex networks.
  Examples are the degree distribution $P(k)$, average degree $\langle k \rangle$, 
clustering coefficient $C$, the ``betweenness'' of a vertex $i$ and ``assortative mixing'' 
describing correlations among vertices in the network.

  Nowadays, an important research issue within complex network (graph) field is the study and 
identification of its community structure, a problem also known as graph partitioning.
  Many definitions of community are presented in the literature.
  In essence, this amounts to divide the network into groups where vertices
inside each group share denser connections among them when compared with
connections across any two groups.
  The main concerns in proposing methods to find communities are in 
developing well successful automatic discovery computer algorithms and  
execution time that can not be prohibitive for large network sizes $n$. 

  More recently various methods have been proposed to find good divisions of networks
\cite{NewmanEPJB04,Arenas05}.
  In particular, some techniques are based on Betweenness measures
\cite{NewmanPRE69}, resistor network \cite{HubermanEPJB38}, Laplacian eigenvalues 
\cite{NewmanXXX06,Munoz04}, 
implementing quantitative definitions of community structures in networks \cite{Loreto04} 
or through out benefit functions known as modularity \cite{NewmanPRE69,NewmanXXX06}.
  Those methods discover communities in time runs that typically scale with the
network size as ${\cal{O}} (n^3)$ or even  ${\cal{O}}(n^4)$.
  However, there is a proposal that scales linearly in time 
but needs a parameter dependent considerations \cite{HubermanEPJB38}. 
  This method views the network as an electric circuit with current flowing throught 
all edges represented by resistors.
  The automatic community finding procedure is hampered by the need of electing
two nodes (poles) that lie in different communities and defining a threshold in
voltage spectrum.

  Here we show how random walkers on graphs, also in connection with electrical networks, 
unveil the hierarchies of subnetworks or the so called community structure.
  Our method combines Laplacian eigenvalue approach with electrical network theory. 
  A brief review of how the spectral graph theory can characterize the structural 
properties of graphs using the eigenvectors of the Laplacian matrix, related to the 
adjacency matrix, has been presented by Newman \cite{NewmanXXX06}.

  The main aspect of the method relies on a generalization of the usual transition
probability matrix ${\bf P}$. 
  The matrix element $P_{ij}$ means the probability for a walk on a weighted graph
at $i$ to its adjacent vertex $j$.
  The interpretation of conductances, the inverse of resistances, among any vertices 
leads to the definition of an effective transition matrix that accounts for hops
on the graph.
  Defining a similarity matrix as a function of the effective transition matrix elements 
it is possible to extract a topological feature related to the manner graph $G$ has 
been organized.
  It turns out that this topological feature corresponds to hierarchical
classes of vertices which we interpret as communities of the network theory.

  To explain our method, we present the essential of the spectral analysis of 
Laplacian matrices in Section 2. 
  In Section 3 we present the arguments leading to the similarity matrix that sets
a scale to extract the community structure.
  In Section 4 we describe how to implement the algorithm and show the
results for the karate club network studied by 
Zachary \cite{Zachary} and for the model designed by Ravasz and Barab\'asi 
\cite{Barabasi03}, an example of network with scale-free property and modular
structure.
   Section 5 concentrates our discussions on weighted graphs and  the final Section 6 
contains our conclusions.

%%%%%%%%%%%%%%%%%%%%%%%%%%%%%%%%%%%%%%%%%%%%%%%%%%%%%%%%%%%
\section{Laplacian eigenvalues and transition matrix}
\indent
  Let us consider a {\it simple} graph $G$, i.e., undirected and with no loops or 
multiple edges, on a finite vertex set $V= \{1, 2, \cdots , n\}$ and edge set $E$, 
represented by the adjacency matrix ${\bf A}$.
  The degree $k_i$ for each vertex $i$ is obtained from the adjacency matrix 
${\bf A}$ as $k_{i} = \sum_{j=1}^n A_{ij}$.
  For non-weighted graphs, the symmetric $n \times n$ adjacency matrix takes values 
$A_{ij}=1$, if there is an edge connecting vertices $(i,j)$ and 0 otherwise.
  Thus, $k_i$ counts the number of edges that connect the selected vertex 
$i$ to other vertices. 
  This extends naturally to weighted adjacency matrix but we leave its version to Section 5.

 For our purpose we study the graph $G$ through a positive semidefinite matrix representation.
 This  is  achieved in the usual manner using the Laplacian.
 The Laplacian matrix of a graph $G$ on $n$ vertices, denoted by ${\bf L}(G)$, is 
simply the matrix with elements
\begin{equation}
 L_{ij} = \left\{ \begin{array}{cl}
  k_i~ & ~{\rm if}~~ i=j   \\ 
  -1 ~ & ~{\rm if}~~ i ~{\rm and}~ j~{\rm are~ adjacents} \\
 ~~0 ~ & ~{\rm otherwise} \, , 
\end{array}
\right. 
\end{equation}
which corresponds to the degree diagonal matrix minus the adjacency matrix, ${\bf L = K- A}$.
  The Laplacian matrix has a long history. 
  It was introduced by Kirchhoff in 1847 with a paper related to electrical 
networks \cite{Grone91} and consequently is also known as Kirchhoff matrix.

  The Laplacian matrix is real and symmetric. 
  Moreover, ${\bf L}$ is a positive semidefinite singular matrix with $n$ eigenvalues 
$\lambda_i$ and eigenvectors $v_i$.
  If we label the eigenvalues in increasing order
$ \lambda_1 \leq  \lambda_2 \leq  \cdots \leq  \lambda_n$, we have ${\bf L}(G)\, v_1=0$.
  The eigenvalue  $\lambda_1 = 0$ is always the smallest one  
and has the normalized eigenvector $ v_1 = (1, 1, \cdots , 1)/ \sqrt{n}$.
  Since the matrix ${\bf L}(G)$ is singular, it has no inverse, but in such cases it 
is possible to introduce the so-called generalized inverse 
$({\bf L}^{\dagger})$ of ${\bf L}$ according to Moore and Penrose's 
definition \cite{Campbell_book}. 
 
  Among many properties for the second smallest eigenvalue $\lambda_2(G)$, known as 
the algebraic connectivity, we recall that \cite{Grone91,Kliemann05}
$\lambda_2(G) = 0  ~~{\it iff}~ G~ {\rm is~ not~ connected}$. 
  For connected networks, the eigenvector components of the first non-null 
eigenvalue ($\lambda_2$) has been applied as an approximate method for 
grouping vertices into communities \cite{Hall70,Munoz04,NewmanXXX06}.
  However the success in partitioning depends on how well $\lambda_2$ is separated from other
eigenvalues.
 
  From now on we identify the graph $G=(V,E)$ with an electrical network 
connected by edges of unit resistances \cite{BollobasBook,voltage_probability}.
 A random walk on $G$ is a sequence of states (vertices) chosen among
their adjacent neighbors.
 To describe the overall behavior of a walker on $G$, one needs to go beyond
the usual analysis of Markov chains with transition matrix $P_{ij}$,
probability to go from vertex $i$ to an adjacent vertex $j$, to include also
hops, i.e., moves across the graph.
  For this end, we evaluate the effective resistances $r_{ij}$ between all distinct 
vertices $i$ and $j$ of $G$.
  Those effective resistances $r_{ij}$ can be numerically evaluated by means
of the electrical network theory as \cite{Gutman03,Gutman04}
\begin{equation}
  r_{ij} ~=~ (L^{\dagger})_{ii} + (L^{\dagger})_{jj} - 
             (L^{\dagger})_{ij} - (L^{\dagger})_{ji} \, ,
\end{equation}
for $i \neq j$ and $r_{ij}=0$ for $i=j$.
  Here, ${\bf L}^{\dagger}(G)$ is the Moore-Penrose generalized inverse of 
the Laplacian matrix ${\bf L}(G)$. 
  Its definition amounts to write ${\bf L}^{\dagger}(G)$ as
\begin{equation}
 (L^{\dagger})_{ij} = \sum_{k=1}^{n-1} \frac{1}{\lambda_k}\, v_{ki} v_{kj} \, .
\end{equation}
  This leads to a simple formulation of the effective 
resistances between all pairs of vertices
as a function of the eigenvalues and eigenvectors of ${\bf L}(G)$,
\begin{equation}
  r_{ij} ~=~ \sum_{k=1}^{n-1}\frac{1}{\lambda_k}(v_{ki}- v_{kj})^2 \,.  \label{rij}
\end{equation}
  As a natural generalization, it is convenient to define the effective conductances 
$c_{ij}$ for all pairs of vertices $(i,j)$ as $ c_{ij} = 1/ r_{ij}$, for $i \ne j$.

  As a consequence of the above results it is possible to extend the usual
random process that moves around through adjacent states $i$ and $j$ to hops 
on the graph. 
  We define the hop transition probability from vertex $i$ to any vertex $j$ by 
\begin{equation}
  P_{ij} ~=~ \frac{c_{ij}}{c_i} \, ,                          \label{Pij}
\end{equation}
where $c_{ij}$ is the effective conductance from $i$ to $j$ and $c_i = \sum_j c_{ij}$. 
  Since a connected network is considered, the probability that a walker who
begins the run at any given vertex $i$ and reaches any other given vertex 
does not vanish.

%%%%%%%%%%%%%%%%%%%%%%%%%%%%%%%%%%%%%%%%%%%%%%%%%%%%%%%%%%%
\section{METHOD}
\indent
 Although $P_{ij}$ is not necessarily equal to $P_{ji}$, it is possible to 
describe hierarchical classes of states perceived by the walker as follows.
 
 Firstly, we consider the generalized ``distance'' expression,
\begin{equation}
  d^{(q)}_{ij} ~=~  \frac{\left( 
        \sum_{k \neq i,j}^{n} |P_{ik}- P_{jk}|^q \right)^{1/q}}{n-2} \,,  \label{dij}
\end{equation}
where $q$ is a positive real number, as a similarity measure between any
vertices.
  Small $d^{(q)}_{ij}$ would imply high similarity between $i$ and $j$ and could be used
to set a hierarchical classification.
  Unfortunately this measure does not provide a good score to classify those 
states into communities.
  We have realized that the fluctuations $S_{ij}$ in $|P_{ik}- P_{jk}|$ indeed play 
the main role for that classification.
  Let us take $q=1$ and define
\begin{equation}
  \overline{d}_{ij} ~=~ \frac{\sum_{k \neq i,j}^{n} |P_{ik}- P_{jk}|}{n-2}     \label{dbar}
\end{equation}
as the average ``distance'' between $i$ and $j$. The standard deviation between those
vertices is given by
\begin{equation}
  S_{ij} ~=~ \left[ \frac{1}{n-3} \sum_{k \neq i,j}^{n} 
             \left( |P_{ik}- P_{jk}|-\overline{d}(i,j) \right)^2 \right]^{1/2} 
                                                           \,.                       \label{Sij}
\end{equation}
  As a matter of fact, this quantity gives a better description of the similarity 
among the vertices in opposite to the average value in Eq. (\ref{dij}). 
  The importance of those fluctuations to classify vertices into communities
may be surmised saying that we should not ask how far away two vertices are, but who are 
their neighbors.
 
  Secondly, we explore the behavior of $P_{ij}$ because low transition probability 
to go from state $i$ to $j$ means that state $j$ is less accessible from state $i$. 
  On the other hand, high transition probability among states defines
a class of easily connected states.
  This is better understood in terms of $1/P_{ij}$. Since the elements
$P_{ij}$ are not necessarily symmetric, we define how close $i$ and $j$ 
are by taken as distance 
${\rm min} \{1/P_{ij}, 1/P_{ji} \} = 1/{\rm max} \{P_{ij}, P_{ji}\}
\equiv 1/P^{max}_{\{ij\}}$.
  In other words, the quantity $1/P^{max}_{\{ij\}}$
sets different levels of transient classes on $G(V,E)$.

  Thirdly, in order to have a well defined class of states we should expect small
transition probability for leaving it.
  Let us also introduce the notation 
$P^{min}_{\{ij\}} \equiv {\rm min}\{P_{ij}, P_{ji}\}$. 
 Thus, a large value of $\Delta_{ij} \equiv P^{max}_{\{ij\}} - P^{min}_{\{ij\}}$
is consequence of small value for the leaving probability
$P^{min}_{\{ij\}}$ and large value for $P^{max}_{\{ij\}}$.

  Therefore, we extract the desired hierarchical analysis  defining heuristically 
a similarity matrix (or ``distance matrix'') 
${\bf D}$ taken simultaneously into account the above remarks:
\begin{equation}
  D_{ij} ~=~ S_{ij}\, 
     \frac{ {\rm max} \{\Delta_{ij}, P^{min}_{\{ij\}}\}} { P^{max}_{\{ij\}}} \,.  \label{Dij}
\end{equation}
   Comparative values of $P^{min}_{\{ij\}}$, for different $(i,j)$ pairs, may be translated 
as a penalty when they are rather large, which has an intimate connection with $\Delta_{ij}$.
   Thus, the maximum between $\Delta_{ij}$ and  $P^{min}_{\{ij\}}$ 
enters in the nominator of Eq. (\ref{Dij}) as an extra 
term to help to set a similarity (or proximity) scale. 
   As we will show in the next sections, the symmetric matrix ${\bf D}$ is able to unveil 
the entire transient classes of states.

%%%%%%%%%%%%%%%%%%%%%%%%%%%%%%%%%%%%%%%%%%%%%%%%%%%%%%%%%%%
\section{Evaluating community identification}
\indent
   To understand the meaning of those transient classes we investigate
in some examples the structure of $G(V,E)$ encoded by the similarity matrix.
  Our analysis reveal well-defined classes of vertices. 
  They occur at different levels of the hierarchical tree under $D_{ij}$ with the 
interesting interpretation of communities i.e., with the structure of well-defined subnetworks.

\subsection{Performance on artificial community graphs}
\indent

 Before discussing a particular issue on how to implement the algorithm
we report its performance on graphs with a well known fixed community
structure \cite{NewmanPRE69}.
 Our method was tested on large number of graphs with $n=128$ vertices
and designed to have four communities of 32 vertices. 
 Each graph is randomly generated with  probability $p_{in}$ to connect vertices in the 
same community and probability $p_{out}$ to those vertices in different communities. 
 Those probabilities are evaluated in order to make the average degree of each
vertex equals to 16. The test amounts to evaluate the fraction of vertices
correctly classified as a function of $z_{out}$, the average number of
edges a given vertex has to outside of its own community. 
 Our algorithm classifies correctly vertices into the four communities for small 
values of $z_{out}$, decreasing its performance towards $z_{out}=8$. 
 We have, for example, the fractions 
$0.99\pm 0.01$, $0.95\pm 0.01$, $0.81\pm 0.02$, $0.57\pm 0.03$, 
respectively for  $z_{out}=5, 6, 7$ and 8.
 The error bar was evaluated over 100 randomly generated graphs.
 Those results are competitive with the analyzed  algorithms in Ref. \cite{Arenas05}. 
 Moreover, we stress that the proposed method is fully parameter independent.
 Also, its computational cost is limited to methods in computing the eigenvalues
and eigenvectors of symmetric matrices. In general it amounts to initial ${\cal O}(n^3)$
operations, with subsequent less expensive iterations ${\cal O}(n^2)$.

\subsection{A graph with leaves}
\indent
  The method is quite simple and much of the computer time is spent in calculating
the eigenvalues and eigenvectors of $\bf L$.
  All that remains to calculate is the effective resistances in Eq. (\ref{rij})
and, with the elements $P_{ij}$, the final similarity matrix $\bf D$ in Eq. (\ref{Dij}).
  However, some care is needed when the graph presents what we call leaves.
  This is explained as follows.
 
  We present in Fig. 1 a small graph to display the information contained
in the matrix ${\bf D}$ and how to perform the hierarchical analysis.
  This example shows a graph containing a subgraph with tree-like topology.
  A tree is a connected acyclic graph.
  In this example, the tree is the subgraph with vertex numbers 5, 6 and 7, which we 
call leaves.
  Their effective resistances are $ r_{56}=r_{57}=r_{35}=1$ and therefore we have 
$ r_{36}=r_{37}=2$.
  For tree-like subgraphs the effective resistances correspond to
the number of edges $\ell_{ij}$ connecting vertices $i$ and $j$.
Therefore, $r_{ij}= \ell_{ij}$ for acyclic branches.
  Also $r_{48}=1$ because there is only one way of reaching vertex 8 from vertex 4. 
  On the other hand, whenever we have different paths joining adjacent vertices $(i,j)$, 
we obtain $r_{ij} < 1$ as consequence of calculating the effective resistance of resistors
connected in parallel and in series.
 For example, $ r_{89}=r_{8(10)}=r_{9(10)}=0.6667$.
 To unveil the hierarchical structure of graphs with leaves, we need to proceed as 
follows because well-defined transient classes of states are only identified for graphs 
with no local tree-like topology.
  Suppose we start with a graph with $m$ vertices ($m=10$). 
  If the graph has leaves, we collect leaf after leaf to remove acyclic 
branches and we end up with a reduced number of vertices $n$ ($ n < m$).
  After collecting all leaves, we work with the Laplacian matrix of order $n$
obtained from the reduced adjacency matrix.
  During this process we keep trace of the original labels. 
  The hierarchical structure of this example is shown in Fig. 2 as a dendrogram
where we have joined the previously removed vertices (6, 7 and 5) to vertex 3
because they naturally belong to the same community as vertex 3 does.
  All presented dendrograms have their similarity (y-axis) ${\bf D}$ scaled to be in
the range $(0, 100)$. This allows a comparative display of their branches.

%%%%%%%%%%%%%%%%%%%%%%%%%%%%%%%%%%%%%%%%%%%%%%%%%%%%%%%
%%%%%%%%%%     example of simple network N=10     %%%%%%%%%
%%%%%%%%%%%%%%%%%%%%%%%%%%%%%%%%%%%%%%%%%%%%%%%%%%%%%%%
%FIGURE 1
\begin{figure}[t]
\begin{center}
\begin{minipage}[t]{0.45\textwidth}
\centering
\includegraphics[angle=0,width=0.65\textwidth]{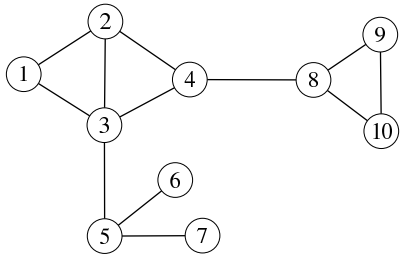}
\caption{A simple graph with a tree-like subgraph: vertices 5, 6 and 7.
 Our graph figures are drawn using VISONE (www.visone.de).}
\label{fig1}
\end{minipage}
\end{center}
\end{figure}

%FIGURE 2
\begin{figure}[h]
\vskip1cm 
\begin{center}
\begin{minipage}[t]{0.45\textwidth}
\centering
\includegraphics[angle=0,width=0.65\textwidth]{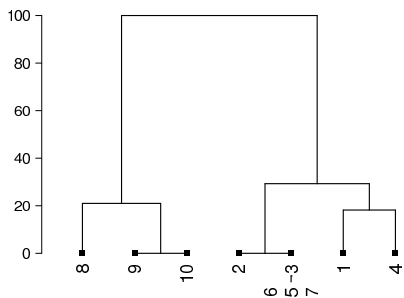}
\caption{The community structure of graph in Fig. 1 is depicted 
as a hierarchical tree or dendrogram with the complete linkage method
for the similarity matrix $D$.
 Our dendrogram figures are drawn with the data plotting package and 
 programming language R (http://www.R-project.org). }
\label{fig2}
\end{minipage}
\end{center}
\end{figure}
%%%%%%%%%%%%%%%%%%%%%%%%%%%%%%%%%%%%%%%%%%%%%%%%%%%%%%%%%%%%%%%%%%%%

\subsection{Zachary karate club network}
\indent
  To illustrate further the meaning of transient classes on $G(V,E)$ from global 
information carried out by ${\bf D}$ we analyze two well known networks in the literature.

 The first example (Fig. 3) corresponds to the network of members of the 
karate club studied by Zachary \cite{Zachary}.
 This graph contains a single leaf: member 12. 
 Our analysis led to the hierarchical structure shown in Fig. 4 by means of
a hierarchical clustering tree, defining communities at different levels. 
 The two main communities reproduce exactly the observed splitting of the Zachary club 
and studied by different community finding techniques 
\cite{NewmanPRE69,NewmanEPJB04,Arenas05,HubermanEPJB38,Munoz04,
Loreto04,NewmanPRE6904,Zhou_hier}.
  Interestingly, a smaller community presented by the hierarchical tree can be clearly 
identified in Fig. 3. It consists of members displayed with shaded circles.
  This small group is only influenced by its members and has a direct interaction with
the instructor.

%%%%%%%%%%%%%%%%%%%%%%%%%%%%%%%%%%%%%%%%%%%%%%%%%%%%%%%
%%%%%%%%%%     karate club network N=34     %%%%%%%%%
%%%%%%%%%%%%%%%%%%%%%%%%%%%%%%%%%%%%%%%%%%%%%%%%%%%%%%%
%FIGURE 3
\begin{figure}[t]
\begin{center}
\begin{minipage}[t]{0.45\textwidth}
\centering
\includegraphics[angle=0,width=0.90\textwidth]{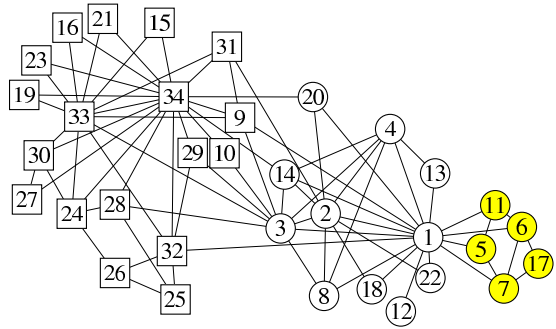}
\caption{The karate club network studied by Zachary. Individual numbers represent
the members of the club and edges their relationships as observed outside the 
normal activities of the club. Squares and circles indicate the observed final 
splitting of the karate club into two communities led by the administrator (34) 
and the instructor (1). A clear further splitting is identified with shaded circles.}
\label{fig3}
\end{minipage}
\end{center}
\end{figure}

%FIGURE 4
\begin{figure}[h]
\vskip1cm 
\begin{center}
\begin{minipage}[t]{0.45\textwidth}
\centering
\includegraphics[angle=0,width=0.85\textwidth]{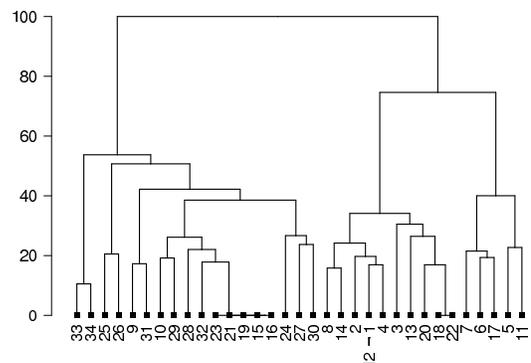}
\caption{The hierarchical structure of network in Fig. 3 is shown
as a dendrogram with the complete linkage method. It correctly identifies the two 
main communities of the karate club.}
\label{fig4}
\end{minipage}
\end{center}
\end{figure}

\cleardoublepage
\newpage
%%%%%%%%%%%%%%%%%%%%%%%%%%%%%%%%%%%%%%%%%%%%%%%%%%%%%%%%%%%%%%%%%%

\subsection{Ravasz and Barab\'asi square hierarchical network}
\indent
  The second example is shown in Fig. 5. It was designed by Ravasz 
and Barab\'asi \cite{Barabasi03} as a prototype of hierarchical organization
we may encounter in real network with scale-free topology and high modularity.
 The main figure is built with the module in (a). A similar figure but with more 
connections between vertices can be built with the module in (b). 
  The study of $D_{ij}$ reveals community structures at different hierarchical levels in Fig. 6, 
respectively for the graphs generated with the modules (a) and (b).

  The hierarchical trees present similar structures, but the hierarchical levels in both 
figures clearly display different network formation patterns.
  Moreover, the hierarchical formation pattern of $G(V,E)$ with branches at different 
heights may be seen as a measure of how cohesive those subgroups are.
  The normalized scale for $D_{ij}$ then can be used to also set degrees of
cohesiveness related to the community formation.

%%%%%%%%%%%%%%%%%%%%%%%%%%%%%%%%%%%%%%%%%%%%%%%%%%%%%%%%%%%
%%%%%%%%%%     hierarchical square by Barabasi N=25   %%%%%
%%%%%%%%%%%%%%%%%%%%%%%%%%%%%%%%%%%%%%%%%%%%%%%%%%%%%%%%%%%
%FIGURE 5
\begin{figure}[t]
\begin{center}
\begin{minipage}[t]{0.45\textwidth}
\centering
\includegraphics[angle=0,width=0.85\textwidth]{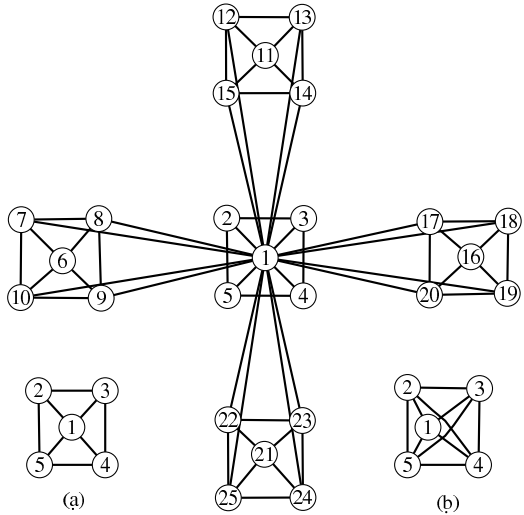}
\caption{The deterministic hierarchical scale-free model with $n=5$ vertices
proposed by Ravasz {\it et al.} \cite{Barabasi02}. 
It is built by generating replicas of the small 5-vertex module (a) shown at left side.}
\label{fig5}
\end{minipage}
\end{center}
\end{figure}

%FIGURE 6
\begin{figure}[t]
\begin{center}
\begin{minipage}[t]{0.45\textwidth}
\centering
\includegraphics[angle=0,width=0.95\textwidth]{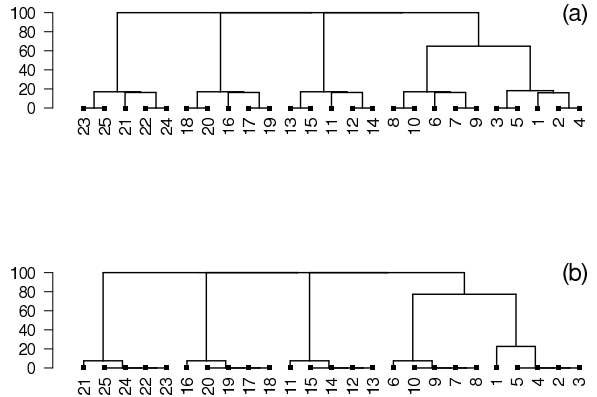}
\caption{Hierarchical structure for the formation pattern of 
         the network in Fig. 5. Dendrogram (a) refers to the network built with module (a)
in Fig. 5 whereas dendrogram (b) refers to the graph built with module (b) in Fig.5.}
\label{fig6}
\end{minipage}
\end{center}
\end{figure}

%%%%%%%%%%%%%%%%%%%%%%%%%%%%%%%%%%%%%%%%%%%%%%%%%%%%%%%%%%%%%%%%%%%
\section{WEIGHTS ON THE EDGES}
\indent
 Our method also applies to graphs such that each edge has a positive real number, the
weight of the edge. 
 The structure of the graph is now represented by the corresponding weighted adjacency 
matrix $\bf W$. It assigns weight $w_{ij} > 0$ if and only if $i$ and $j$ are
connected vertices and 0, otherwise. 
 The concept of the Laplacian matrix extends directly to weighted edges,
${\bf L}(G)=  {\bf E}(G)-  {\bf W}(G)$, where $E_{ii}=\sum_{j=1}^{n} w_{ij}$
is the diagonal weighted matrix whose values are the total weight of the edges
adjacents to vertex $i$.
 Again, ${\bf L}(G)$ is a real symmetric matrix where the row sums and the column sums are
all zero. Thus, we have the same spectral properties as recalled to the particular case
$w_{ij}=1$ for all adjacent vertices $i$ and $j$.
  Therefore, the method presented to unweighted graphs extends naturally to weighted ones
with no change in the algorithm.

\subsection{Performance on artificial community weighted graphs}
\indent
 We have also verified the performance of this method on weighted graphs with fixed
community structure \cite{NewmanE70W}. Our test is performed on the same artificial
graphs randomly generated as described in Section 4.A.
 The computer generated graphs have 128 vertices and are divided into four groups
of 32 vertices. 
 Here, edges among vertices are randomly chosen such that the average degree is fixed at 16.
 The test is performed for the most difficult situation where $z_{out}=z_{in}=8$.
 That is, each vertex has as many adjacent connections to inside as to outside its community. 
 For each graph, we attach a weight $w > 1$ to the edges inside each community 
and keep the fixed weight 1 for those edges which lie between communities.
 We evaluate again the fraction of vertices classified correctly as a function of $w$.
 As $w$ increases from the starting value 1, the weights enhance the community structure.
 This is clearly highlighted by our method.
 Our performance amounts to the following fractions of correctly classified vertices, 
$0.89, 0.94, 0.97$ and 0.98, respectively for $w=1.4, 1.6, 1.8$ and 2. 
 The averages were calculated over 100 randomly generated graphs, with 
error bars smaller than 0.01.

\subsection{identifying cohesive subgroups}
\indent
 As an example, we apply our method to the problem of analyzing weighted interactions 
related to verify how pairs of teachers are engaged in professional discussions
\cite{Frank96}.
  This is a social network with $n=24$ members. Their edges are characterized by the 
professional discussions in a high school, called ``Our Hamilton High'', during the 
1992-1993 school year. Teachers were asked to list and weight the frequency of their 
discussions in that school to at most five teachers. 
 This way of attributing weights leads to a directed network.
 The weights should follow a scale running from 1, for discussions occuring less than 
once a month, to the largest weight value 4, for almost daily discussions \cite{Frank96}.
  Every vertex number contains characteristics of teachers as gender, race, subject field,
room assignment, among others.
  To perform our analysis we have defined the weights to each edge
as the average of the values placed on the edges in the original directed network.
Thus, this new weighted network is characterized by edges with real values in the range 
$(0.5, 4)$ as representing the interactions among the members of that school.
  The community structure revealed by our analysis is represented by the dendrogram in Fig. 7.
  Its structure exhibits the formation of various communities. For comparison with
the results in \cite{Frank96}, we also pick out the four main groups. 
  The study of the their members reveals an association mainly according to race and gender,
as also found in Ref. \cite{Frank96}. However, there are some differences in 
the members identification in each group. This may be due to the fact we are not
analyzing exactly the same weighted network: our network is made undirect throught out
an average process while the original one was handled in its original directed form.

%%%%%%%%%%%%%%%%%%%%%%%%%%%%%%%%%%%%%%%%%%%%%%%%%%%%%%%
%%%%%%%%%%     teacher_weight_Frank N=24     %%%%%%%%%
%%%%%%%%%%%%%%%%%%%%%%%%%%%%%%%%%%%%%%%%%%%%%%%%%%%%%%%
%FIGURE 7
\begin{figure}[t]
\begin{center}
\begin{minipage}[t]{0.45\textwidth}
\centering
\includegraphics[angle=0,width=0.90\textwidth]{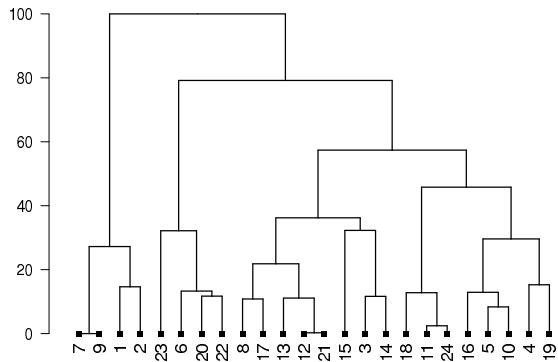}
\caption{Network community of professional discussions among teachers at 
``Our Hamilton High''.}
\label{fig7}
\end{minipage}
\end{center}
\end{figure}
%%%%%%%%%%%%%%%%%%%%%%%%%%%%%%%%%%%%%%%%%%%%%%%%%%%%%%%%%%%%%%%%%%%%%%%%%

%%%%%%%%%%%%%%%%%%%%%%%%%%%%%%%%%%%%%%%%%%%%%%%%%%%%%%%%%%%%%%%%%%%
\section{Conclusions}
\indent
 In conclusion, random walks on graphs in connection with electrical
networks highlight a topological property of $G(V,E)$: transient classes of 
vertices which we interpret as communities in the original graph.
 Here we emphasize that those special classes of vertices are a direct consequence of 
effective transition probabilities, which display a global perspective 
about the map of interactions that characterize the graph.
 We demonstrate its high performance in identifying community structures
in some examples which became benchmark for initial algorithm validation.
 Moreover, it is parameter tunning independent.
 Our criterion to define communities depends only on $G(V,E)$ and not on any
explicit definition of what a community structure must be.

  It is likely that our proposed algorithm may produce new insights for large
graphs. Application examples may include protein-protein interactions and
the compartment identification in food-web structures. 
  The visual information about how members form communities along the
hierarchical tree may permit understand and characterize cohesive communities.

{\bf Acknowledgments} \\

The author acknowledges valuable discussions with  O. Kinouchi, A.S. Martinez
 and the support from the Brazilian agencies CNPq (303446/2002-1) and FAPESP (2005/04067-6). 
%%%%%%%%%%%%%%%%%%%%%%%%%%%%%%%%%%%%%%%%%%%%%%%%%%%%%%%%%%%%%%%%%%%

%%%\section{References}

%%%%%%%%%%%%%%%%%%%%%%%%%%%%%%%%%%%%%%%%%%%%%%%%%%%%%%%%%%%%%%%%%%%%%%%%

\end{document}